\documentclass[traditabstract]{aa}

\makeatletter

\usepackage{pdflscape}
\makeatother

\usepackage[usenames, dvipsnames]{color} 
\usepackage{color}
    \usepackage{textcomp}
\begin{document}

\title{Polyacenes and diffuse interstellar bands} 

\author{A. Omont\inst{1}
\and H.\ F.\ Bettinger\inst{2}
\and  C.\ T\"{o}nshoff\inst{2}}
\institute{Sorbonne Universit\'e, UPMC Universit\'e Paris 6 and CNRS, UMR 7095, Institut d'Astrophysique de Paris, France
\and  Institut für Organische Chemie, Universität Tübingen, Auf der Morgenstelle 18, 72076 Tübingen, Germany}

\abstract{
The identification of the carriers of the diffuse interstellar bands (DIBs) remains to be established, with the exception of five bands attributed to C$_{60}^+$, although it is generally agreed that DIB carriers should be large carbon-based molecules (with $\sim$10-100 atoms) in the gas phase, such as polycyclic aromatic hydrocarbons (PAHs), long carbon chains or fullerenes. The aim of this paper is to investigate more specific possible carriers among PAHs, namely elongated molecules, which could explain a correlation between the DIB wavelength and the apparent UV resilience of their carriers. More specifically, we address the case of 
polyacenes, C$_{\rm 4N+2}$H{$_{\rm 2N+4}$}, with N$\sim$10-18  fused rectilinear aligned hexagons. Polyacenes are attractive DIB carrier candidates because their high symmetry and large linear size allow them to form regular series of bands in the visible range with strengths larger than most other PAHs, as confirmed by recent laboratory results up to undecacene (C$_{46}$H{$_{26}$}).
Those with very strong bands in the DIB spectral domain are just at the limit of stability against UV photodissociation.  
They are part of the prominent PAH family of interstellar carbon compounds, meaning that only $\sim$10$^{-5}$ of the total PAH abundance is enough to account for a medium-strength  DIB.
After summarizing the limited current knowledge about the complex properties of  polyacenes and recent laboratory results, the likelihood that they might meet the criteria for being carriers of some DIBs is addressed by reviewing the following properties:  wavelength and strength of their series of visible bands; interstellar stability and abundances, charge state and hydrogenation; and DIB rotation profiles.
 No definite inconsistency has been identified that precludes polyacenes from being the carriers of some DIBs with medium or weak strength, including  the so-called C$_2$ DIBs. 
But, despite their many interesting properties, additional experimental data about long acenes and their visible bands are needed to make robust conclusions.
}


\maketitle

\section{Introduction}

\noindent Since their discovery almost one century ago the origin of the diffuse interstellar bands (DIBs)
has remained an outstanding problem (see, e.g., 
Herbig 1995; Sarre 2006, Tielens \& Snow 1995; and especially Cami \& Cox 2014; see also, e.g.,\ Omont, 2016; Fan et al.\ 2017; Cami et al.\ 2018 for more recent references).
Today, the number of  
bands confirmed as DIBs  has reached almost 500 (see e.g.,\ Hobbs et al.\ 2009, 2008). The
strongest bands are  conspicuous and 
ubiquitous, not only along lines of sight in the Galactic disk, but also  in other galaxies. 

While the carriers of only five DIBs (in the near infrared) have been identified  as C$_{60}^+$ (Foing \& Ehrenfreund 1994; Campbell et al.\ 2015, 2016a,b; Walker et al.\ 2015, 2016, 2017; Campbell \& Maier 2017a, 2018; Spieler et al.\ 2017; Lallement et al.\ 2018), they are still the only DIBs whose carriers are identified. However, there is more or less general agreement  that DIB carriers should be large carbon-based molecules (with $\ga$10 to $\sim$100 atoms) in the gas phase (e.g.,\ Tielens 2014),  although other possibilities remain open  (see e.g.,\ Snow 1995, 2014). 

In addition to a precise and consistent wavelength  match across the whole spectrum, the most stringent constraint on such possible DIB carriers comes from the strength (equivalent width) of the bands, which can be expressed as the product of the molecular (M) carrier abundance  ${\rm \chi_{M}}$ by the band strength  (e.g., Cami 2014; Omont 2016). For medium-strength DIBs with equivalent width per unit reddening, EW\,$\sim$\,20 {\rm m\AA /mag}, it may be written as

\begin{equation}
{\rm X_{CM}~x~f \approx 2\,x\,10^{-6} ~~or~~ \chi_{M}~x~f \approx 10^{-10}}
,\end{equation}

\noindent where f is the oscillator strength 
and X$_{{\rm CM}}$ denotes the fraction of total interstellar carbon locked up in the  
molecule M with N$_{\rm C}$\,$\sim$\,60 carbon atoms. This implies for example ${\rm X_{CM}}$ $\sim$ 10$^{-4}$  for f $\sim$ 2 x 10$^{-2}$
or  ${\rm X_{CM}}$ $\sim$ 10$^{-6}$  for f $\sim$ 2.

 The most popular candidates for DIB carriers are therefore found among the  three main allotropic forms of  known interstellar carbonaceous particles: linear carbon chains, polycyclic aromatic hydrocarbons (PAHs), and fullerene-like compounds. Despite intensive spectroscopic efforts, all attempts to identify DIBs with known spectra of long carbon chains and PAHs have been unsuccessful up to now (see, e.g., Campbell \& Maier 2017a; Zack \& Maier 2014; Salama et al.\ 2011; Salama \& Ehrenfreund 2014). Although it remains possible that the three varieties of carbon compounds could include many DIB carriers, all three suffer from various difficulties. 

For example, most fullerene compounds seem to lack the required combination of strong band strength and large interstellar abundance (Cami 2014; Omont 2016).  
For most of them, f-values of the visible bands are at most a few times 10$^{-2}$, although those with endohedral metals may well have high
f-values as discussed by Kroto and Jura (1992), 
 and the abundance of the most obvious compound, C$_{60}^+$, is only a few times 10$^{-4}$ of interstellar carbon. But various types of  molecules, such as fullerene cages with endohedral metals, various  C$_{60}$ adducts, and open cages, cannot be excluded.

Carbon-chain derivatives, which are known in dense molecular clouds with C$_n$ skeleton and n\,$\la$\,10, and in diffuse clouds as C$_2$ and C$_3$,  have been repeatedly proposed as DIB carriers (see, e.g.,  Douglas 1977; Thaddeus 1995; Snow 1995; Allamandola et al.\ 1999; Maier et al.\ 2004; Zack \& Maier 2014). Despite questions about their formation and interstellar stability 
and the exclusion  of short chains (n$\la$10) from the lack of coincidence of their known wavelengths with those of DIBs (e.g., Zack \& Maier 2014),  they remain good candidates for DIB carriers. Their spectral properties, including a series of strong bands throughout the visible range for which both $\lambda$ and f-values can be roughly proportional to the number n of carbon atoms, have made them especially appealing as DIB-carrier candidates (e.g., Campbell \& Maier 2017a).
But there is still no evidence of chains long enough to be stable in the diffuse interstellar medium (ISM), that is,\ with n$\ga$25. Furthermore, the proposed processes for their formation from grain shattering (e.g., Duley 2000; Jones 2016) remain tentative.

On the other hand, the prominent mid-IR emission bands of PAHs (L\'eger \& Puget, 1984; Allamandola et al.\ 1985) show that PAHs as a whole contain several percent (up to $\ga$10\%) of the total interstellar carbon (e.g., Puget \& L\'eger 1989; Tielens 2008, 2013). 
After their identification as carriers of the mid-IR bands, it was a logical step to propose that PAHs are also major DIB carriers  (van der Zwet \& Allamandola 1985; L\'eger \&  d’Hendecourt 1985; Crawford, Tielens \& Allamandola 1985; Salama et al.\ 1996). 
However, to this day all attempts to identify a single specific interstellar PAH have failed (Salama \& Ehrenfreund 2014). Only interstellar detection of two aromatic molecules with a single six-membered ring, benzene (C$_6$H$_6$) and benzonitrile (C$_6$H$_5$CN), was achieved by Cernicharo et al.\ (2001) and McGuire et al.\ (2018), respectively.
 In view of the failure to identify individual PAHs and  the expected chemical variety for interstellar PAHs, it seems unlikely that any randomly  selected PAH could be abundant enough to be responsible for a strong DIB (e.g., Steglich et al.\ 2011).
 
However, from their detailed discussion of the proposal that the PAHs might be the main carriers of the DIBs, Salama et al.\ (1996) concluded that this proposal seems extremely promising, provided that "the interstellar PAH distribution is dominated by a small, finite number of species (100-200)", especially in the form of cations.  Among the immense variety in size, isomers, hydrogenation, and ionization of PAHs, it is generally believed that the most abundant ones are the most compact ("pericondensed") with a number of carbon atoms large enough to be stable  against photodissociation, such as circumcoronene  C$_{54}$H$_{18}$. These latter are often referred to as 'grandPAHs' (e.g., Tielens 2013; Andrews et al.\ 2015). However, such large PAHs remain difficult to study and most gas-phase PAH spectroscopic studies have been limited to a number of carbon atoms N$_{\rm C}$\,$\la$20-25 (e.g.,  Gredel et al.\ 2011; Salama et al.\ 2011; Huisken et al.\ 2014; Hardy et al.\ 2017; Kofman et al.\ 2017) with few  exceptions (Steglich et al.\ 2011; Campbell \& Maier 2017b; Zhen et al.\ 2016, 2018),  thereby hampering the search for any coincidence with known DIBs. 
It could be that most of the well-studied PAHs are too small to survive photodissociation in the diffuse ISM, while the variety of larger PAHs  and their isomers and the lack of symmetry of most of them distribute their global visible interstellar absorption among many weak bands (e.g., Cocchi et al.\ 2014). 

A peculiar class of PAHs has perhaps been overlooked as potential DIB carriers, namely long "catacondensed" PAHs (but see, e.g., Ruiterkamp  et al.\ 2002, 2005; Halasinski et al.\ 2003). 
In particular, linear acenes or polyacenes C$_{\rm 4N+2}$H$_{\rm 2N+4}$ (with N fused  hexagons on a straight line; Fig.\ 1) may appear especially promising. Their structure can be seen as two coupled polyene chains, which makes them somewhat similar to "long carbon chains", explaining that their optical spectrum also displays a series of strong bands, shifted into the visible range for N $\ga$ 10 (e.g., Ruiterkamp  et al.\ 2005; Malloci et al.\ 2011; Bettinger et al.\ 2016).
Recent measurements (Bettinger et al.\ 2016; Shen et al.\ 2018; see §2.3, Fig.\ 2) show the existence of very strong bands (f $\sim$ 3-4) in the range $\sim$4000-5000\,\AA\ for N\,=\,9 and 11, and  provide very precise values of their wavelength in argon matrix.  Similar strong bands of higher acenes, from N\,=\,12 to $\sim$16-18, should cover the whole main DIB spectral range $\sim$5500-7000\,\AA .

In addition to their high symmetry and strong bands, other important features of polyacenes could point to good candidates as carriers of some DIBs, especially those sensitive to UV destruction of the 'C$_2$' and '$\zeta$'    
 ~families (Sect. 4.2).
Their interstellar photodissociation could match the varied behaviour of these DIBs versus extinction (Sect. 4.2). 
The large value of one of their rotation constants  could explain DIB profiles (Sect. 4.3).
But the higher reactivity and instability of polyacenes compared to other PAHs might make  their interstellar formation difficult (Sect. 2.4). As other PAHs, they could be formed by shattering of carbon interstellar grains or larger PAHs, but specific proof is lacking.

In this paper, we review the properties of polyacenes and present a series of arguments showing that they are potentially interesting carriers of some  DIBs. The paper is organized as follows. Section 2 summarizes the known properties of small acenes and presents what can be inferred from our understanding of longer, more complex acenes.  
Section 3 discusses the expected properties of polyacenes in the diffuse ISM: charge, hydrogenation, photodissociation, possible formation, and abundance. 
Section 4 addresses
the likelihood of polyacenes meeting the various criteria for DIB carriers: 
wavelength and strength of their visible bands; interstellar abundances, charge and hydrogenation; DIB rotation profiles. 
Finally, Sect. 5 summarizes the reasons why polyacenes could be potential DIB carriers and provides suggestions for future theoretical and experimental work to further explore this possibility.

\section{Summary of polyacene properties}

\subsection{Overall and electronic structure}
Acenes or polyacenes are the most threadlike PAHs made up of N linearly fused benzene rings, C$_{\rm 4N+2}$H{$_{\rm 2N+4}$ (Fig.\ 1).
With their derivatives they are a typical organic semiconductor material that is being intensively studied (e.g., Anthony 2008; Ye \& Chi 2014). However, beyond pentacene they become extremely reactive (Sect. 2.5) and unstable, which explains why their synthesis was achieved only very recently, especially for octacene, nonacene, decacene, and undecacene 
(Bettinger \& T\"{o}nshoff 2015; Zuzak et al.\ 2017; Kr{\"u}ger et al.\ 2017;  
Shen et al.\ 2018; Zuzak et al.\ 2018), and not beyond yet. 


\begin{figure}[htbp]
         \begin{center}
\includegraphics[scale=0.6, angle=0]{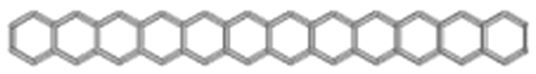}
 \caption{Example of a polyacene,  dodecacene (N=12), C$_{50}$H$_{28}$.  
In the normal hydrogenated form there is an H atom bound to each outer C atom.
}
     \end{center}
 \end{figure}

In some way, polyacenes are the most extreme  catacondensed   ~PAHs, the opposite of pericondensed ~compact ones. More characteristically, they are among the most symmetric PAHs, with D$_{\rm 2h}$ point group  symmetry. 
However, distortion is possible by out-of-plane deformation. 
Their rotation structure is that of a very elongated, near-prolate, slightly asymmetric top. For instance, the rotation constants of pentacene are (Heinecke et al.\ 1998): A\,=\,44.5, B\,=\,3.9 and C\,=\,3.6.\,x\,10$^{-3}$cm$^{-1}$. For higher acenes, A-values scale roughly as N$^{-1}$ and B and C as N$^{-3}$, meaning that the ratio A/(B or C) is $\sim$\,50-100. Table C.1 gives the values of the three rotational constants for N\,=\,9-20 derived from density functional computations using the B3LYP hybrid functional in the spin-unrestricted formalism (UB3LYP) and the 6-31G(d) basis set as implemented in Gaussian 09
(see Becke 1993; Lee, et al.\ 1988; Frisch et al.\ 2009). For instance, for N\,=\,14, rotational constants A\,=\,1.62x10$^{-2}$, B\,=\,2.16x10$^{-4}$ and C\,=\,2.13x10$^{-4}$\,cm$^{-1}$  were computed. 

Such high symmetry and elongation make  their PAH electronic structure peculiar. Although their aromaticity is limited to a single Kekul\'e hexagon, the delocalisation of the latter makes the conjugation of the $\pi$ electrons extremely extended. It may result in very peculiar structures of the electronic distribution including biradical singlet states (see, e.g., Zade \& Bendikov 2012 and references therein) and  even higher polyradical character, depending on the length, as was predicted by theory (e.g., Hachmann et al.\ 2007; Plasser et al.\ 2013). 
A characteristic feature is the low gap between the highest  occupied molecular orbital (HOMO) and the lowest unoccupied molecular orbital (LUMO).
Similarly, the singlet-triplet energy splitting decreases to less than 1 eV for large N.

Their complexity contributes to the uncertainty about the visible
spectra of polyacenes beyond those which have been actually observed (Bettinger et
al.\  2016, Shen et al.\ 2018, see Sect. 2.2). 
In any case, the acenes should keep the very rich structure of electronic and vibronic excited states of other large carbonaceous molecules with a comparable
number of carbon atoms N$_{\rm C}$, resulting in a high probability  the excited states have a very
short lifetime and efficient internal energy
conversion (IEC); see Sect. 2.2.

As noted by Ruiterkamp et al.\ (2005), polyacene ionization potentials are significantly smaller than those of pericondensed PAHs (from $\sim$6.5\,eV for pentacene and hexacene, $\sim$5\,eV for N\,$\sim$\,10-20, to a limit of probably $\la$4.5\,eV for very large values of N; Wu et al.\ 2016).  This could make  their photoionization by interstellar UV radiation easier.
As for long 'carbon chains' and other PAHs,  high values of the electron affinities of acenes (1.4\,eV, $\sim$2-2.3\,eV,  $\sim$2.5-2.8\,eV for N\,=\,5, 10, 20, respectively, see, e.g., Malloci et al.\ 2011; Chai 2014; Wu et al.\ 2016) should favor anions in interstellar regions that are somewhat UV shielded. 

While acenes are the most extreme and symmetric cases, various other catacondensed PAHs
may be considered as DIB-carrier candidates, as done in general by Ruiterkamp et al.\ (2005) who pointed out their possibly large f-values. 
Such molecules, 
 especially of the broader acene family, share many of the polyacene properties (e.g., polyphenacenes and other similar elongated isomers of polyacenes, Luzanov et al.\ 2017; pyrene-fused acenes, Li et al.\ 2016; azethrenes, Huang et al.\ 2016; azaacenes, Li \& Zhang 2015). However, we limit our discussion to pure acenes because other catacondensed PAHs do not appear to be as good candidates as acenes for carriers of strong DIBs\footnote{Other compounds, such as oligorylenes, have been proposed as attractive candidates for DIB carriers (Ruiterkamp et al.\ 2002; Halasinski et al.\ 2003); but see Ruiterkamp et al.\ (2005).}. 
The same is true for hetero-acenes where, for example, at least one CH group is replaced by an N atom, because of their lower symmetry and lower expected interstellar abundances. 

Together with their linear form,  the cyclic form of acenes, [N]cyclacenes, C$_{\rm 4N}$H{$_{\rm 2N}$, might be present in the interstellar medium }
(Jones 2016). Although they have not yet been synthesized because of their instability, 
theoretical studies of their electronic properties (Wu et al.\ 2016, Battaglia et al.\ 2017) have confirmed that the properties of cyclacenes with an even number of benzene rings are similar to linear acenes.  
Similar elementary belts, such as [12]cyclophenacene, 
 whose carbon skeletons may also be regarded as constituents of carbon nanotubes (CNTs), have recently been synthesized (Povie et al.\ 2017, 2018 and references therein).  


\begin{figure*}[htbp]
         \begin{center}
\includegraphics[scale=0.6, angle=0]{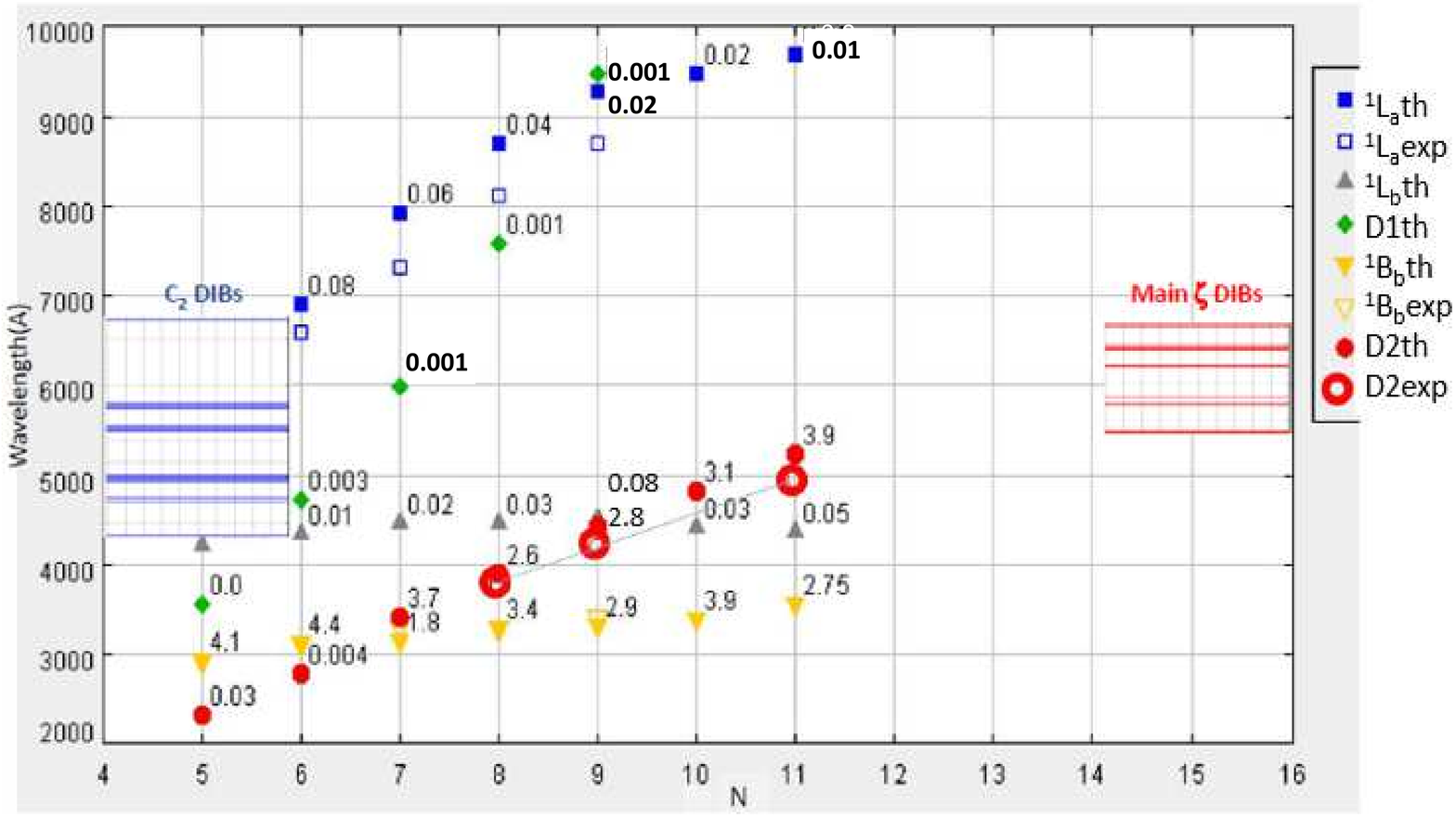}
 \caption{Wavelengths of optical bands of higher acenes with N hexagons from 
Bettinger et al.\ (2016) and Shen et al.\ (2018): {\it full symbols}, theoretical values with f-values; 
{\it open symbols}, experimental Ar-matrix data  taken  from  T{\"o}nshoff \& Bettinger (2010), except for undecacene (N\,=\,11; Bettinger in preparation).
 See the symbol explanation in the right column, with the names of the excited molecular levels and the labels  'exp' for measured wavelengths and 'th' for theoretical estimates.
See text (Sect. 4.3) for estimates of order of magnitude of matrix shifts. 
The short horizontal blue lines on the left show the wavelength of the   C$_2$ DIBs (Thornburn et al. 2003), and the main $\zeta$ DIBs (Fig.\ 4) are marked by short horizontal red lines on the right.
The expected extrapolation to longer wavelengths  shows that similar strong bands of higher polyacenes, N$\sim$12-16, should extend through the whole main spectral domain of DIBs from $\sim$5000\,\AA\ to  $\sim$7000\,\AA\ (Fig.\ 3) .}
     \end{center}
 \end{figure*}

\subsection{Spectra and oscillator strengths}
\noindent 

 Despite the uncertainty on the structure of their electronic levels  for N\,$>$\,11 (Sect. 2.1), it is clear that all acenes have a number of important bands in the visible domain, as do other PAHs of similar size. Some of these bands may be much stronger (at least a factor $\sim$3) than visible bands of other PAHs with similar N$_{\rm C}$ (e.g., Malloci et al. 2007)  because of the high symmetry and the elongation of polyacenes.

This can be seen more precisely  from Fig.\ 2 which  displays the wavelengths of optical bands of higher acenes based on argon matrix and theoretical values from Bettinger et al.\ (2016),  including new results for undecacene (N=11) which displays  a strong band (f\,$\sim$\,4) located at 543\,nm  in a matrix of polystyrene (Shen et al.\ (2018) and at 495\,nm in a matrix of argon (Bettinger et al.\ in preparation). 
  For N\,$\le$\,9,  
the strongest bands with f\,$>$\,1 are confined to the UV range. Nonacene has already a strong band  close to 4000\,A, and it is clear from Fig.\ 2 that, despite the uncertainties, the trend toward redder wavelengths will bring at least a strong band into the visible range for N\,$\ge$\,10 probably up to N\,$\sim$16-20. This means that   a series  of strong bands of neutral acenes with f\,$\ga$\,1 is expected in the main DIB spectral range from $\sim$4500\,\AA\  to $\sim$7000\,\AA . 
 However,  the complexity of the behaviour of electronic excited states of polyacenes does not allow one to extrapolate the gas-phase position of these strong visible bands for N\,$\ge$\,12 with the precision required for the identification of DIB carriers (Fig.\ 2). 

 In addition, relatively strong visible bands should also exist for polyacene cations and anions, as  has been measured up to nonacene (Mondal et al. 2009; T{\"o}nshoff \& Bettinger 2010). 
 Strong infrared bands are also expected for ions (Mondal et al.\ 2009; Malloci et al.\ 2007)  and   weaker ones for neutrals.

\subsection{Internal energy conversion and fluorescence}

Similarly to other PAHs,  strong internal energy
conversion  is expected among electronic excited states of polyacenes (especially the highest ones even above the ionization limit; e.g., Leach 1995; Zhen et al.\ 2015) and their vibronic components. For neutral acenes, after absorption of a near- or far-UV photon, IEC may end with the usual pattern where the energy is shared between fluorescence (and possible phosphorescence) and mostly conversion to pure  vibrational energy through transit to a triplet state.  For many PAHs, fluorescence may be either direct emission from a low-lying excited state, or delayed after transit through a triplet state of slightly lower energy ('Poincar\'e' fluorescence, L\'eger et al.\ 1988; see Ji et al.\ 2017 for its evidence in anthracene). But Poincar\'e fluorescence seems unlikely for higher polyacenes, since 
the extrapolated optical gap is $\sim$1\,eV (Shen et al.\ 2018), while the triplet state is computed to be only a few kilocalories per mole above the ground singlet state. Therefore,  a significant population of the first bright singlet state from the triplet state would be very unlikely due to the $\sim$1\,eV energy difference. 

Fluorescence is known to be strong in the case of pentacene  
 but weak for hexacene (Nijegorodov et al.\ 1997). 
The  fluorescence properties of higher acenes remain unknown. They should depend on the detailed IEC transfer between excited singlet and triplet states. However, it is possible that fluorescence is strong for high-N neutral acenes due  to the very high f-values of some visible bands (Fig.\ 2). On the other hand, this seems less likely for ions whose level structure is dominated by doublet states.

\subsection{Reactivity}
\noindent 

Long acenes are known for their extreme reactivity which makes them unstable and has yet prevented their laboratory study  beyond N\,=\,11. As quoted, for example by Zade \& Bendikov (2012),  the reactivity of the smaller acenes increases with their length, especially on the central carbon rings, and a greater tendency towards a biradical structure is observed (see also Yang et al.\ 2016). 

We note that all the rings of polyacenes are susceptible to reaction with a carbon atom by inserting it  into a peripheral six-membered ring. Through this process,  a seven-membered  carbon ring is formed as suggested by experiments and quantum chemical computations by Krasnokutski et al.\  (2017){\footnote{See also similar reactions between atomic carbon and benzene, Kaiser et al.\ (1999, 2003); Hahndorf et al.\ (2002).}}. This possibility is more reduced for pericondensed PAHs such as circumcoronene because C cannot be inserted into a CC bond common for two aromatic rings.
 As with other PAHs, polyacenes should easily react with atomic oxygen. The most important outcome should be ejection of CO, either directly or after intermediate formation of an epoxy.

\smallskip
\section{Expected properties of polyacenes in the diffuse interstellar medium}

\subsection{Photodissociation}

Polyacenes should be present in the diffuse 
ISM just as the multitude of other PAHs, and share most of their properties, especially those of other catacondensed PAHs. 
Photodissociation is the main cause of processing interstellar PAHs. Its modelling has been intensively discussed in the last 30 years, as summarized in Appendix A for the diffuse ISM, with the following main conclusions. The H-loss photolysis may result from single UV-photon absorption for medium-size PAHs (N$_{\rm C}$\,$\la$\,40, N$_{\rm (C+H)}$\,$\la$\,55) and 
 should proceed first before C-fragment ejection from the carbon skeleton. 
Because of the roughly exponential dependence of the photodissociation rates on the activation energy  E$_{\rm A}$ for photoinduced dissociation and  the vibrational temperature, interstellar PAHs should be  either mostly normally hydrogenated or mostly totally dehydrogenetated. 

For polyacenes, as discussed in Appendix A, there should thus exist a critical value of N, N$_{\rm cr}$, around which they rapidly pass from fully hydrogenated to fully dehydrogenated in normal diffuse clouds. However, it is possible that for N very close to N$_{\rm cr}$, partially dehydrogenated polyacenes may exist, that is,\ including the coexistence of several hydrogenation states for the same carbon skeleton. 
Quantitatively,  it is  roughly  expected that such partially dehydrogenated polyacenes might be the dominant form  in the diffuse ISM for perhaps N\,$\sim$11 or 12.  Fully hydrogenated polyacenes could then be found with N\,$\ga$\,12-13 in the diffuse ISM, and with N\,$\ga$\,11-12 in 'C$_2$' translucent clouds, but these values remain uncertain.

The  carbon skeleton should be stable in most of this  range of N values for normal hydrogenated polyacenes. But this appears questionable for smaller N values, with N$_{\rm C}$\,$<$\,40, and even for higher N values, in all cases when  full dehydrogenation (carbon nanoribbons) is expected. In the latter case, it is probable that the acene linear carbon skeleton rapidly transforms into a more stable isomer that could be either more compact (graphene nanosheets, also known as graphene quantum dots), or include carbon pentagons, heptagons, or larger rings (Mackie et al.\ 2015; Parneix et al.\ 2017). 
In addition, as other PAHs, polyacenes should be prone to erosion and eventual destruction, either by chemical reactions, in particular with atomic oxygen, or by more energetic processes than single-photon absorption, such as two-photon absorption (Montillaud et al.\ 2013; Andrews et al.\  2016) and carbon-atom knock-out or disintegration by energetic particles (Micelotta et al.\ 2010; Chabot et al.\ 2017). 
However, because of their likely greater efficiency in accreting C$^+$ (Sect. 2.4),  the decay of their carbon skeleton might be slowed compared to pericondensed PAHs. With partial dehydrogenation, this could be important by successive edge accretions of interstellar C$^+$ ions.

\smallskip
\subsection{Charge state}

Like other PAHs, it is expected that interstellar polyacenes  exist in various charge states -- neutral, cation, and anion -- in the diffuse ISM. But the ratio of cations to neutrals should be higher than for  pericondensed PAHs since the ionization potential of acenes is significantly lower (Sect. 2.1).
However, the  expected  ionization fractions are at least as uncertain as those derived in the various models for general PAHs (see, e.g., Le Page et al.\ 2001, 2003; Montillaud et al.\ 2013; Andrews et al.\ 2016), mostly because of the uncertainty in the electronic recombination rates. On the other hand, due to the high values of the electron affinities of neutral acenes, the importance of the polyacene anions should not be overlooked, especially in translucent regions that are UV shielded.

\smallskip
\subsection{Possible formation, growth, and abundances}
Although polyacenes present exceptional symmetry and properties, 
their actual interstellar abundance remains uncertain. Because of the general ignorance about the generation of interstellar PAHs and their shape and isomer distribution, one may only propose  qualitative arguments about the role of polyacenes. In the frame of the most likely top-down generation processes of PAHs from disruption of carbonaceous grains and large PAHs (e.g., Scott \& Duley 1997; Duley 2000; Jones 2014; Chiar et al.\ 2013; Zhen et al.\ 2014), it is often proposed that thread-like carbonaceous compounds might be favored 
(Scott \& Duley 1997; Duley 2000; Chabot et al.\ 2017). 
Elongated partially aromatic compounds may be produced as a result of the expected structure of the interstellar  hydrogenated carbon grains (e.g., Pendleton \& Allamandola 2002). 
Further interstellar processing of such chain-like compounds (UV ablation and isomerization, C$^+$ accretion, etc.) might therefore  favor elongated PAHs, but their instability should penalize polyacenes.
For instance, polyacenes should be less thermodynamically stable than their phenacene isomers, since for example the heat produced by the formation of anthracene is more positive (53 kcal/mol) than that of phenanthrene (48 kcal/mol).

More generally, it is  believed that the general interstellar processing of all PAHs  favors the most stable PAHs in pericondensed shapes, especially a small number of so-called  grandPAHs   ~(e.g., Andrews et al.\ 2015). This might well rule out any significant interstellar abundance of polyacenes. 
However, in the absence of any individual identification of interstellar PAHs and of their actual isomer distribution,  we may also consider the  above arguments in favor of polyacenes.
The growth of the  polyacene carbon skeleton might perhaps proceed by successive edge accretions of interstellar C$^+$ ions onto the edge of dehydrogenated forms.  
Similar reactions of partially dehydrogenated PAHs  with C$_2$H$_{\rm x}$ or C$_4$H$_{\rm x}$ have been proposed for the observed PAH formation on nanoparticles of silicate or SiC (Tian et et al.\ 2012; Zhao et al.\ 2016).

In principle,  the variety of possible alterations of polyacenes should also be considered, such as 
 superhydrogenation (with several H atoms bound to the same edge C atom), methylation, C addition,  hetero atoms, sputtering by atomic oxygen, ring opening and other isomerization, and even perhaps cyclization from partially dehydrogenated polyacenes. Of course,  most of these peculiarities would disqualify them as carriers of most DIBs.  But it may be expected that they cannot survive for long in the harsh UV environment of the diffuse ISM which could favor the evolution into elongated PAHs. 

\smallskip
\section{Could  polyacenes meet criteria for DIB carriers?}
In this section, we address the evidence for and against polyacenes being significant carriers of some DIBs.
Aspects that are reviewed include the wavelength and strength of visible bands of polyacenes, their interstellar abundance, charge, and hydrogenation 
 and finally their expected interstellar band profiles. 

\subsection{Wavelengths, strengths, and ionization of visible bands}

From Fig.\ 2, it is striking to see how the  series of the expected strongest visible bands of the 
 polyacenes   
from N\,$\sim$\,11 to $\sim$16-18 matches the main DIB wavelength domain ($\sim$4000-7500\,\AA ). 

It is interesting to note that the required amount of carbon locked in a given polyacene carrier of an average DIB would remain quite modest. Assuming f\,$\sim$\,1, which seems  warranted for the first values of N\,$\ge$\,11 (Fig.\ 2), 
and a medium-strength DIB with EW\,=\,20\,m\AA/mag, Eq.\ (1) yields X$_{\rm C}$\,$\approx$ 2\,10$^{-6}$ for the fraction of total interstellar carbon locked in such an acene. This is only about  2\,10$^{-5}$ of the total carbon amount in PAHs. 

The various  ionization states of  polyacenes might also partly account for the different 'families' of DIBs (see, e.g., Vos et al.\ 2011; Thorburn et al.\ 2003; Fan et al.\ 2017), although  the strength of the visible bands  of N$\ga$11 acene ions could be smaller than for neutrals, and} the expected ionization percentages remain uncertain as for all PAHs because of unknown electron recombination rates (Sect. 3.2).  
The 'C$_2$' DIBs, which need the shielded environment of translucent clouds (Thorburn et al.\ 2003; Ka{\'z}mierczak et al.\ 2010; Elyajouri  et al.\ 2018, and references therein), might be carried by relatively small neutral polyacenes (Sect. 4.2) (or by anions).  Ruiterkamp et al.\ (2005) considered such a possibility for the  'C$_2$' DIBs observed on the translucent line of sight of  HD\,147889. Also note that most heavy ionized acenes have strong near-IR bands (Malloci et al.\ 2007), meaning that they might account for various near-IR DIBs, together with visible DIBs.

 \subsection {The dependence of  C$_2$  and $\zeta$ DIBs on $\lambda$}


\begin{figure}[htbp]
         \begin{center}
\includegraphics[scale=0.3, angle=0]{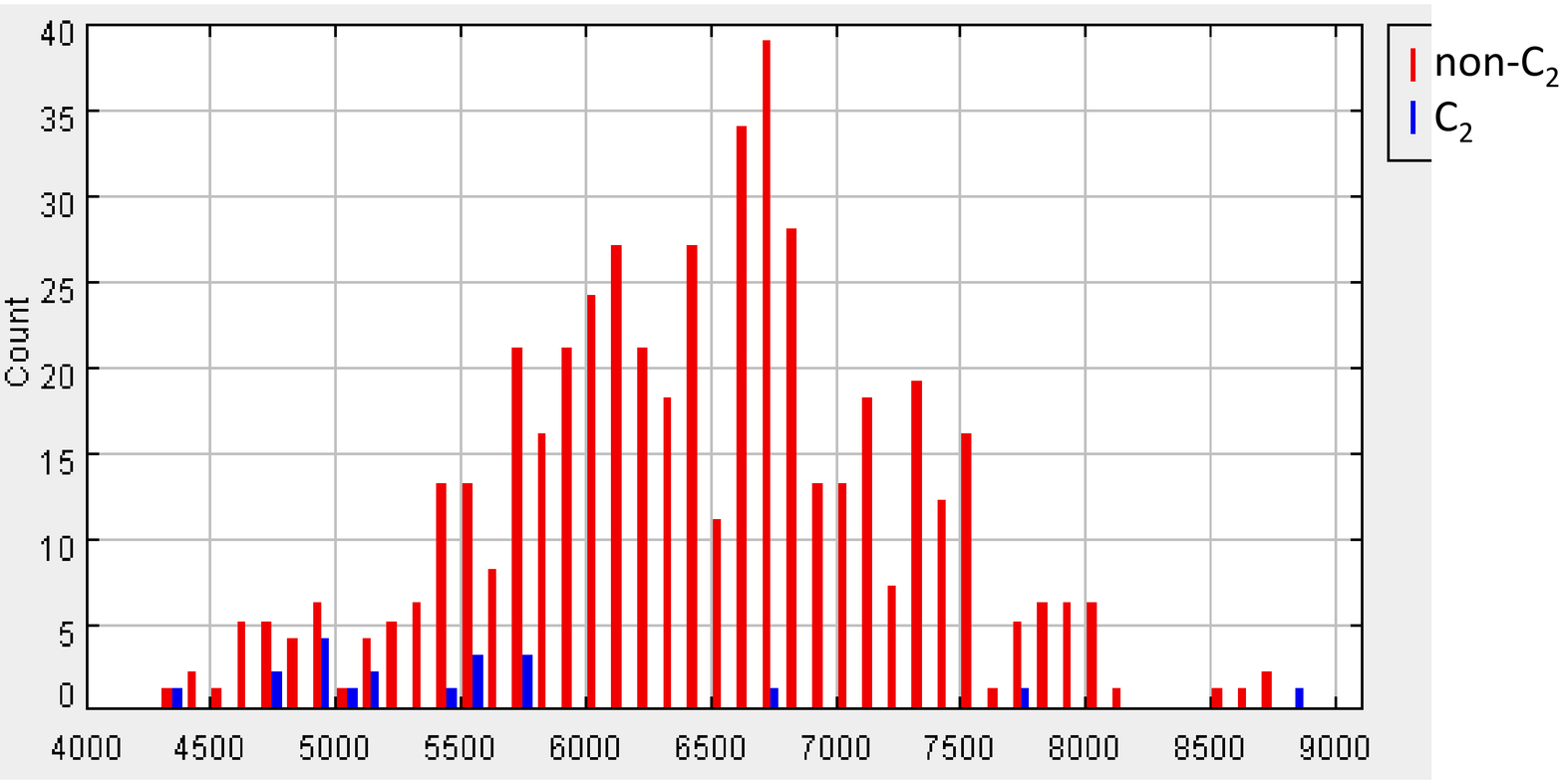}
 \caption{Wavelength distribution of C$_2$ DIBs from Thorburn et al.\ (2003) and Cami et al.\ (2018) compared to the other DIBs from Hobbs et al.\ (2009).}
     \end{center}
 \end{figure}


\begin{figure}[htbp]
         \begin{center}
\includegraphics[scale=0.34, angle=0]{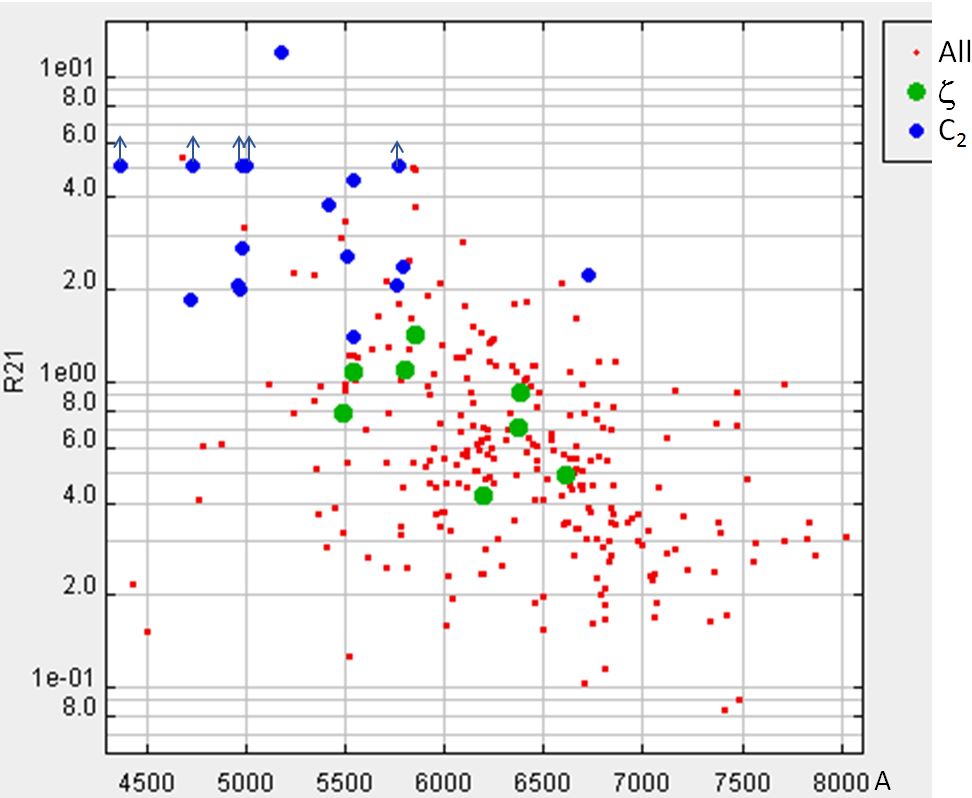}
 \caption{Wavelength dependence of the ratio R$_{21}$ of the equivalent widths of the  C$_2$ DIBs (4364, 4727, 4735, 4964, 4969, 4980, 4985, 5004, 5176, 5419, 5513, 5542, 5546, 5763, 5769, 5793, 6729\,\AA , Thornburn et al.\ 2003) and of the main $\zeta$ DIBs (5494, 5545,  5797,  5850,  6196,  6376,  6379,  6614\,\AA )  listed by Ensor et al.\ (2017), on the lines of sight of  HD\,204827 and  HD\,183143 (Hobbs et al. 2008, 2009). The vertical arrows indicate a lower limit of this ratio for the  C$_2$ DIBs not detected on the line of sight of  HD\,183143. Small red dots denote all other 260 DIBs detected on both lines of sight  (Hobbs et al. 2008, 2009).}
     \end{center}
 \end{figure}

A characteristic feature of the series of strong optical bands of polyacenes  either observed for N\,$\le$\,11 or expected for higher N values is the strong dependence of their wavelength on the length of the molecule (see Fig.\ 2 and also Ruiterkamp et al.\ 2005). Maybe this size dependence could be correlated with the  DIB-wavelength dependence of the fragility to UV exposure of the $\zeta$ and C$_2$ DIB carriers.
It is well known that the most fragile DIB carriers are those of the C$_2$ DIBs which need UV-shielded environments to survive (e.g., Thorburn et al.\ 2003), while these DIBs have shorter average wavelengths than others. This is seen in  
Fig. 3 which displays the wavelength distribution of the C$_2$ DIBs  compared with the other DIBs.  It is striking that all C$_2$ DIBs but three are confined below 5800\,\AA .

This $\lambda$ dependence of the sensitivity to UV may be somewhat extended to other DIBs as a whole. 
One may roughly characterize the DIB fragility to UV exposure by the ratio of their equivalent widths, R$_{21}$,
 in the UV-shielded line of sight HD\,204827 (Hobbs et al.\ 2008) to the unshielded one HD\,183143 (Hobbs et al.\ 2009). 
The wavelength dependence of this ratio is displayed in Fig.\ 4 for all the DIBs detected in both sightlines (260) 
  and it is compared with the  C$_2$ DIBs listed by Thorburn et al.\ (2003). The general decrease of R$_{21}$  with $\lambda$ is clear.  It may be inferred that the average UV resilience of the DIB carriers increases with $\lambda$.

In addition, there are 168 confirmed DIBs on the line of sight of HD\,183143 which are not detected in HD\,204827 (2/3 at $\lambda$\,$>$\,6500\,\AA ), implying R$_{21}$\,$\la$\,1; similarly, 107 DIBs on the line of sight of  HD\,204827 are not detected in HD\,183143 (85\% at $\lambda$\,$<$\,6200\,\AA ), implying R$_{21}$\,$\ga$\,1.  This behavior is consistent with the trends of the R$_{21}$/$\lambda$ correlation displayed in Fig.\ 4.

 A possible explanation is that $\lambda$ strongly depends on the number of carbon atoms N$_{\rm C}$ of these carriers, since we expect an exponential dependence of the H photolysis rate on the number of vibration modes (Appendix A) and hence on N$_{\rm C}$. Such a dependence with N$_{\rm C}$ of the wavelength of strong bands is especially marked for elongated molecules such as carbon chains (e.g., Campbell \& Maier 2017a), elongated catacondensed PAHs (Ruiterkamp et al.\ 2005), including polyacenes (Fig.\ 2). We may infer that elongated molecules play a significant role among DIB carriers. However, other origins for the correlation of R$_{21}$  with $\lambda$ are possible, such as the predominance of cationic bands at large $\lambda$.

As seen in Fig.\ 4, the correlation between R$_{21}$ or the resilience with $\lambda$ also holds in particular for the main $\zeta$ DIBs listed by Ensor et al.\ (2017), whose best definition is the correlation with the H$_2$ column density (see, e.g., Lan et al.\ 2015; Ensor et al.\ 2017). This probably extends to the whole  family of $\zeta$ DIBs whose other members are not yet as well characterized.    
Such a $\lambda$ behavior of the C$_2$ and $\zeta$  DIB families and their carriers, which should be neutral, might be compatible with the $\lambda$ dependence of the strong visible D$_2$ band of polyacenes (Fig.\ 2). As regards the main $\lambda$ range of C$_2$ bands, $\sim$4700-5700\,\AA, first, it is seen in Fig.\ 2 that it corresponds roughly to the gas-phase wavelength of the D$_2$ band of neutral polyacenes with N\,$\sim$\,11-12\footnote{Polyacene anions might also be considered as carriers of the C$_2$ DIBs. However,  the same $\lambda$ specificity favoring shorter wavelengths is not expected as it is for neutral  acenes.}. 
As discussed in Sect. 3 and Appendix A, these acenes with N\,$\sim$\,11-12 could be partially dehydrogenated in the normal diffuse ISM, but might be normally rehydrogenated in the UV-shielded C$_2$ clouds.  

For instance, we may wonder whether  the strong band of undecacene observed at 4950\,\AA\ in an argon matrix (Fig.\ 2 and §2.2) corresponds to a  C$_2$  DIB. This requires an estimate of the argon-matrix shift of the undecacene band. An approximate value for this shift could be its value for the 5362.8\,\AA\ band of pentacene, 779\,cm$^{-1}$ (Halasinski et al.\ 2000; Salama et al.\ 2011). This would yield a gas-phase wavelength of the order of 4755\,\AA\ for undecacene. Taking into account the uncertainty on the matrix shift, this wavelength could be compatible either with the very strong and blended C$_2$ DIB at 4727\,\AA\  or with a weak C$_2$ DIB at 4734\,\AA . Matching the strong  part of the blended DIB at 4727\,\AA\ seems unlikely because of the lack of a similar DIB counterpart for dodecacene of which  the D$_2$ band is expected around 5100\,\AA . A match with the C$_2$ DIB at 4734\,\AA\  or the weak part of the blended DIB at 4727\,\AA\ seems possible, but it lacks significance because of the weakness of the DIBs and the wavelength uncertainty. Similarly, a coincidence of the D$_2$ band of dodecacene is not excluded with one of the C$_2$ DIBs in the range $\sim$5000-5100\,\AA\ (and perhaps of the D$_2$ band of decacene with the C$_2$ DIB at 4364\,\AA ).

If we next consider  the $\zeta$ DIBs in Fig.\ 4, their main $\lambda$ range,  $\sim$5500-6500\,\AA, corresponds roughly to the gas-phase wavelength of the D$_2$ band of neutral polyacenes with N\,$\sim$\,13 to 16-18 (Fig.\ 2). Such a size range might be expected for the smallest acenes that could be stable against H photodissociation in the normal diffuse interstellar medium, with a decreasing stability from N=16 to 13 although the actual photodissociation behavior of polyacenes in this N range remains uncertain  (Sect. 3.1 and Appendix A). If it seems difficult
that such polyacenes can account for the strongest $\zeta$ DIBs, it is not impossible that they carry some of the numerous weaker $\zeta$ DIBs. 

Let us stress again however that such a correlation between the wavelength and the carrier stability against UV photodissociation through the whole DIB spectral domain is not unique to polyacenes. It could also occur for other kinds of  elongated carriers, such as long carbon chains or other classes of elongated PAHs, which have strong visible bands  whose $\lambda$ increases with molecular length. 

 \subsection{Band profiles}

It is clear that the profiles of the DIBs and their variation in different environments contain key information about the nature of their carriers by putting strong constraints on the parameters of the molecular transitions involved. 
Going on various pieces of evidence, there is a general agreement that  the  observed profiles of DIBs result from a combination of several processes, mainly: i) Lorentz broadening, corresponding to an extremely short (ps) lifetime of the excited state because of rapid internal conversion ii) rotational broadening (see, e.g., review by Sarre 2014 and references therein) and iii) vibrational broadening from hot bands  (Marshall et al.\ 2015; Oka et al.\ 2013). To summarize, the broadest profiles must be dominated by IEC, including the most extreme case of the 4428\,\AA\ DIB (Snow et al.\ 2002); the narrow $\zeta$ profiles and their resolved fine structure seem well accounted for by pure rotational broadening (e.g., Cami et al.\ 2004, Sarre 2014 and references therein), as well as the similar profiles of the C$_2$ DIBs (Elyajouri et al.\ 2018), while many $\sigma$ DIBs seem to need a combination of several processes,  including the C$_{60}^+$ bands (Campbell et al.\ 2016a,b, 2018). 

It is recognized that a key constraint on the DIB carrier identification is the variations of their width with the environment and especially the radiation field. These variations are  particularly pronounced for the line of sight towards the star Herschel 36 (He 36).
The  increased width is  generally interpreted as an enhancement of the rotational temperature mostly induced by the strength of local UV and/or IR radiation field.

It has been shown, especially by Oka et al.\  (2013) Huang \& Oka (2015), that such a large broadening cannot be achieved by the small rotational constants of large molecules, such as heavy pericondensed PAHs, fullerenes or very long carbon chains. These authors have thus proposed that the required large rotational constants  imply that the carriers of these DIBs are small molecules with no more than about ten heavy atoms. However, this could be in contradiction with the fragility of such small molecules  to photodissociation in the UV radiation of the diffuse ISM, which explains why they are not observed there. 

However, as noted, for example by Huang \& Oka (2015), a similarly large broadening may be achieved by a nonlinear molecule if only one of its rotational constants A is  large enough (a few 0.01\,cm$^{-1}$). This could be a key advantage for polyacenes (Table C.1) since it is expected for example that A\,$\sim$\,0.016\,cm$^{-1}$ for N\,$\sim$\,14 (§2.1). 

For a transition moment polarized along the long axis of the molecule, as expected for the main D$_2$ band of polyacenes (Fig.\ 2), Appendix C shows that a typical $\zeta$ DIB width of $\sim$1.5-2.5\,cm$^{-1}$ may be achieved with a difference of $\sim$1-2\% between the A rotation constants of the excited and ground states, for a rotational temperature of $\sim$100\,K. 

The prediction of the rotation temperature(s) of interstellar PAHs remains difficult because of the number and complexity of physical processes to be considered. 
Appendix B summarizes previous discussions and their adaptation to the particular case of elongated polyacenes which are close to prolate symmetric tops. 
A dominant process is  the internal exchange of energy between the vibrations and the rotation (IVRET). It  forces the distribution between the K-sublevels of rotation along the main axis of the molecule to be governed by a temperature T$_{\rm K}$ equal to the vibration temperature T$_{\rm vib}$. 
For polyacenes, the combination of IVRET and periodic absorption of UV-visible photons should result in  high values of T$_{\rm K}$ between photon absorptions, perhaps $\ga$100\,K in the diffuse ISM and even  higher values for stronger radiation fields.

 Although all these processes remain  uncertain, it is not impossible that they could explain both the DIB profiles observed in the normal diffuse ISM and their broadening in sightlines with enhanced radiation intensity such as Her 36 (Dalhstrom et al.\ 2013).

\section{Concluding remarks and prospects}

In this paper, we described the characteristics of polyacenes and explored their potential role as carriers of the DIBs. 
In this context, the existence of a possible general correlation of the DIB wavelength with the UV resilience of their  carriers (Fig.\ 4) is worth stressing. It could be an indication of the  important role of elongated molecules among these carriers, since the dependence of $\lambda$ on N$_{\rm C}$  is expected to be significantly  stronger for elongated molecules than for compact ones, as found for polyacenes (Fig.\ 2). 

 Although there is still no decisive argument, the number of indicative clues in support of higher acenes being significant DIB carriers is noteworthy. Polyacenes are members of the PAH family, which includes the most abundant contenders as DIB carriers, and they display unique properties of symmetry and elongation that induce regular series of very strong bands throughout the DIB wavelength domain (Fig.\ 2). It is not excluded that those with very strong  visible bands are just at the limit of stability against UV photodissociation (for N\,$\ga$\,11-14 in the diffuse and translucent ISM).
They should exist in various forms of ionization and degrees of dehydrogenation which might fit with the properties of the different DIB families.
Despite the lower stability of linear polyacenes compared to compact PAHs, some processes might compensate for that in such a way that they reach significant abundances, while only $\sim$10$^{-5}$ of the total PAH abundance is enough to account for a medium-strength DIB.
The characteristic rotational profiles displayed by a number of DIBs and their broadening in strong radiation fields might also be accounted for by special properties of polyacene bands and their large A rotational constant.

All in all, no definite inconsistency has been identified precluding N\,$\sim$\,11-13 polyacenes from being carriers of  main C$_2$ DIBs, including undecacene with its measured strong D$_2$ band, and those with N\,$\ga$\,13-14  being carriers of some  medium-strength DIBs. However, in the absence of precise  laboratory data on the higher acenes and their visible bands, no firm conclusions can be drawn. In addition, the properties of polyacene photodissociation are uncertain. Furthermore,  the mode of formation of interstellar acenes remains largely hypothetical, as is the case for the formation of PAHs in general, and the known higher reactivity and instability of polyacenes compared to other PAHs might make  their interstellar formation difficult. Nevertheless, despite  all these uncertainties, the many interesting properties of polyacenes 
warrant further exploratory work to determine with greater likelihood whether or not they could eventually be  important carriers of the DIBs.

On the experimental side, any eventual acene-carrier identification would need  confirmation from gas-phase or He-droplet spectroscopy, including a careful consistency check of the full spectrum of the putative carrier with the observed spectra (see, e.g., Snow 1995, 2014).
 However, because of the difficulty of such experiments with extremely unstable compounds, the extension of matrix spectroscopy of linear N-acenes  beyond N\,=\,11 is fundamental. 

On the theoretical side, the extension of existing quantum chemical structure modeling 
to higher linear acenes than N\,=\,11 is important for understanding their complex electronic structure and the properties of their visible bands. 
If such results could prove to be encouraging in view of DIB-carrier identification, it would be desirable to extend the computation 
to N-polyacene ions  
and dehydrogenated forms. 
In parallel, it would be important to precisely model the interstellar photodissociation (H atoms first and then the C skeleton)  of acenes.

If the importance of higher acenes as DIB carriers was further supported by experimental and theoretical work,  modelling the profiles of the corresponding DIBs assuming rotational broadening of acene bands would provide important further information. 

Finally, high-quality observations with the best ground- and space-based telescopes will be critical to  further improve the constraints on DIB carriers, and more importantly, once DIBs carriers are identified, to extract all the information they can provide on the physics and properties of the interstellar medium.

\bigskip

\begin{acknowledgements}
We are indebted to Sydney Leach and Benoit Soep for many discussions and important  suggestions, and to Pierre Cox for his careful reading of the manuscript and his suggestions for improving it. 
We would like to thank  Rosine Lallement, Olivier Bern\'e, Christine Joblin, Thomas Pino, Farid Salama, 
Elisabetta Micelottta, etc.\  
for discussions and comments on various aspects. We also thank Madeleine Roux-Merlin for her precious help in finding references.
We thank the referee for his/her helpful detailed suggestions. 
This work was supported in part by the German Research Foundation (DFG). The computations were performed on the BwForCluster JUSTUS. The authors acknowledge support from the state of Baden-Württemberg through bwHPC and the DFG through grant no INST 40/467-1 FUGG.
\end{acknowledgements}

\bigskip

\appendix

\section{Summary of PAH photodissociation in diffuse clouds and the case of polyacenes}

Absorption of UV photons is the most important source for processing interstellar PAHs. For large PAHs, it is expected that most of the $\sim$6-13.6\,eV energy of an absorbed photon is rapidly redistributed among the numerous energy levels of these large molecules, eventually ending in various outcomes -- ionization, various dissociation channels, and isomerization 
-- before cooling by emission of photons mostly in the infrared range. The ionization yield following the absorption of a UV photon remains low, close to the ionization threshold, but  increases considerably for the most energetic interstellar UV photons  (Leach 1995, 1996; Zhen et al.\ 2015, 2016).

Photodissociation is a key process for small and medium-size PAHs. It can result from the absorption of a single UV photon with energy $\sim$10\,eV, especially for releasing H atoms with typical binding energy (BDE) and activation energy, E$_{\rm A}$, in the range $\sim$4.5-5\,eV. The rate of such processes is directly proportional to the UV radiation intensity G$_0$, which is generally expressed in Habing units  (1.6$\times$10$^{-3}$ erg cm$^{-2}$ s$^{-1}$ between 6 and 13.6\,eV). G$_0$ varies mostly between 3 and 10 Habing units in the regions of the diffuse ISM where the DIBs are formed (see, e.g., Fig.\ 18 of Vos et al.\ 2011). 
But it is well established that a two-step processes with the successive absorption of two photons may be dominant for larger binding energy (e.g., Montillaud et al.\ 2013; Andrews et al.\ 2016; Castellanos et al.\ 2018a). It is not impossible that such two-step processes are important when there is a low-energy metastable isomer accessible with a lower activation energy, E$_{\rm A}$\,$\sim$\,4\,eV, and a lower  BDE. Such a case was analyzed in detail by Castellanos et al.\ (2018a) for the perylene cation (C$_{20}$H$_{12}^+$), where the transfer of an H within a trio-H may form an aliphatic CH$_2$ with E$_{\rm A}$\,$\sim$\,3.81\,eV. Such a situation should be similar for all  trio-H  and quarto-H of PAHs, including the quarto-H of polyacenes. 
But it seems unlikely that such metastable isomers can survive until the arrival of a second photon in interstellar conditions.

It is generally thought that the use of simple statistical theories is justified for modeling the photodissociation of interstellar PAHs (Le Page et al.\ 2001;  Montillaud et al.\ 2013; Andrews et al.\ 2016; Castellanos et al.\ 2018b; West et al.\ 2018, and references therein). As expected, the results of these models show a steep exponential variation of the photodissociation rates with activation energy, the size of the PAH, and the energy of the absorbed photon. This makes modeling the photodissociation rates relatively difficult, because varying, for example, an activation energy by a few 0.1\,eV may change a photodissociation rate by several orders of magnitude.   In addition,  it is increasingly accepted that PAH photodissociation is significantly more complicated than assumed in such simplified models. For instance, Parneix et al.\ (2017) show that ring opening and formation of pentagonal rings should occur in parallel or in competition with hydrogen loss, especially for high levels of dehydrogenation. Complete molecular dynamics studies might eventually be required, as performed for example by Simon et al.\ (2018).

In the absence of similar modeling for the actual case of polyacenes, one may try to infer the qualitative behavior of their photodissociation rates from these models for PAHs of similar number of atoms. 
It seems well established that interstellar photodissociation of PAHs proceeds first with release of atomic hydrogen, before producing small C$_{\rm n}$H$_{\rm m}$ molecules with erosion of the carbon skeleton. This should apply to polyacenes as well. 
Quantitatively, for N-polyacenes, we attempt to infer the value of N for which dehydrogenation proceeds in the diffuse ISM (G$_0$$\sim$\,5) from existing  models for other PAHs with similar number of carbon atoms N$_{\rm C}$. For the relevant range of N$\sim$11-14 (Fig.\ 2), i.e.\ N$_{\rm C}$\,$\sim$\,46-58, N$_{\rm (C+H)}$\,$\sim$\,72-84, the case of circumcoronene, C$_{54}$H$_{18}$, seems suitable. It was discussed in detail in particular by Andrews et al.\ (2016) and Castellanos et al.\ (2018a,b).  They show that H-loss for circumcoronene  proceeds at a very low rate, only by two-photon absorption, meaning that the dehydrogenation should be negligible in the diffuse ISM. This should also be the case for N-acenes with  N\,$\ge$\,14 (N$_{\rm C}$\,$\ge$\,58). On the other hand, Castellanos et al.\ (2018a,b) show that ovalene (C$_{32}$H$_{14}$) 
should be completely dehydrogenated.  

Due to the very steep variation of the dehydrogenation rate with N, there should be a  value N$_{\rm cr}$ of N around which the polyacene hydrogenation passes from normally fully hydrogenated to  at least partiallly dehydrogenated for a very small variation of N, $\delta$N\,=\,1 to 3. 
If the polyacenes were found to behave like pericondensed PAHs such as ovalene and circumcoronene with  similar activation energies and isomers, one could infer that the critical value N$_{\rm cr}$ for dehydrogenation should be  about 9 or 10.  However, it is possible that the photodissociation of heavy polyacenes or their cations might be made significantly easier by various peculiarities of these molecules such as the quarto H atoms at the two ends of the molecule, 
the richness of their low-energy electronic levels, 
and especially the possibility 
of nonplanar distortions which might lower some activation energies for isomerization. 
 We  therefore tentatively estimate that it is not even excluded  that N$_{\rm cr}$\,$\sim$\,11-13.

Looking for the detailed properties of the dehydrogenated states of polyacenes, it is generally agreed that the C-H binding energy  is lower from two adjacent vicinal H on a same C-C bond  (in "duo", "tertio" or "quarto" position) than for a single ("solo") H atom at peri positions  (e.g., Andrews et al.\ 2016;  Castellanos et al.\ 2018a). We have verified that by DFT  calculations (B3LYP/6-31G*) for the  quarto  H atoms at the two ends of polyacenes for N\,=\,11 and 13. We find that the "solo-H" have bond dissociation energies 
(E + ZPVE) ranging from 4.8 to 5.3\,eV for N\,=\,11 and 5.1-5.6\,eV for 
N\,=\,13. The quarto H atoms 
have much lower BDE 
of 4.7-4.8\,eV both for N\,=\,11 and N\,=\,13.

In addition, the likely lower activation energy from states with odd number of H atoms might favor partially dehydrogenated states with even H number. It is therefore expected that the first step of dehydrogenation should be the loss of two H at each end of the polyacene molecule, yielding  C$_{\rm 4N+2}$H{$_{\rm 2N}$}. However,  it seems that the loss of four additional  Hs should rapidly follow because of the still low BDE of the remaining quarto H after partial dehydrogenation, as given in general by Castellanos et al.\ (2018a), yielding  C$_{\rm 4N+2}$H{$_{\rm 2N-4}$}.
Such partially dehydrogenated polyacenes might be the dominant form for one or a few values of N in the diffuse ISM, for N in the possible range $\sim$11-12. Fully hydrogenated polyacenes could thus be found with N\,$\ga$\,12-13 in the diffuse ISM, and with N\,$\ga$\,11-12 in 'C$_2$' translucent clouds, but these values remain uncertain.

However, distortion of the hexagonal C-skeleton may arise simultaneously with dehydrogenation, and completely dehydrogenated polyacenes produced by subsequent dehydrogenation seem unstable with formation of carbon pentagons or larger rings (Parneix et al.\ 2017; Mackie et al.\ 2015).

\section{Rotational excitation of interstellar polyacenes}

The prediction of the rotation temperature(s) of interstellar PAHs remains difficult because of the number and the complexity of physical processes to be considered (e.g., Rouan et al.\ 1992, 1997; Ali-Ha{\"i}moud et al.\ 2009, 2013; Ysard \& Verstraete 2010; Hoang et al.\ 2010, 2011; Silsbee et al.\ 2011; Planck Collaboration 2014; and references therein). However, the modeling framework developed in these references may provide basic clues for polyacenes as well, although it was generally designed for the case of disk (i.e.,\ oblate) symmetric-top PAHs, while elongated polyacenes are close to prolate symmetric tops. 

As stated by Rouan et al.\ (1992), the internal exchange of energy between the vibrations and the rotation (IVRET) is a dominant mechanism (see also, e.g., Ali-Ha{\"i}moud 2013). It  forces the distribution between the K-sublevels of rotation along the main axis of the molecule to be governed by a temperature T$_{\rm K}$ equal to the vibration temperature T$_{\rm vib}$. For prolate tops such as polyacenes, the result is indeed different and somewhat simpler than for disk PAHs since, through the large A value, T$_{\rm K}$ determines the main effects of rotation levels on the profiles of parallel visible bands, while the J distribution has only minor effects. Just after the absorption of a UV-visible photon, 
T$_{\rm vib}$ and hence T$_{\rm K}$ are determined by the rate of IR-vibration cooling (Appendix A). However, this phase which lasts as long as  thermal relaxation is achieved through internal vibration redistribution  (IVR) and IVRET, is very brief ($\la$ hours). It  stops when IVR breaks down because the vibration excitation energy is too low, maybe when T$_{\rm vib}$ reaches about 200\,K in the diffuse ISM (e.g., Ysard and Verstraete 2010). Therefore, an interstellar polyacene molecule spends most of its life between photon absorptions ($\sim$0.2 yr)  in an uncertain state where only a few lower vibration levels are populated out of thermal equilibrium. In  parallel, IVRET breaks also down for similar values of T$_{\rm vib}$, probably of at least a couple of hundred Kelvin (Ali-Ha{\"i}moud 2013). Then, T$_{\rm K}$ cannot decay by rotational damping because the permanent electric dipole is zero in such symmetric polyacenes. Despite some possible additional slow cooling through residual IVRET and vibration-IR emission, it is then  likely that T$_{\rm K}$ keeps a high value, perhaps $\ga$100\,K in the diffuse ISM, before the absorption of the next photon, and even a higher value for stronger radiation fields. Although all these processes remain highly uncertain, it is not impossible that they could explain both the DIB profiles observed in the normal diffuse ISM and their broadening in sightlines with enhanced radiation intensity such as He 36 (Dahlstrom et al.\ 2013).

\section{DIB rotational profiles}

\begin{table*}[htbp]
      \caption[]{Calculated rotational constants of polyacenes (UB3LYP/6-31G*)}
         \label{tab:lines}
            \begin{tabular}{ r c c  c c c  c  }
            \hline
           \noalign{\smallskip}
Acene &  A/GHz   & A/cm$^{-1}$  &  B/GHz  &  B/cm$^{-1}$  &  C/GHz  &  C/cm$^{-1}$  \\
      \noalign{\smallskip}
            \hline 
 5Ac$^a$  &   1.335      &   4.45x10$^{-2}$    &        0.117  &              39x10$^{-4}$  &           0.108          &     36x10$^{-4}$  \\ 
 9Ac  &      0.75012      &      2.502x10$^{-2}$  &     0.02291  &          7.642x10$^{-4}$  &   0.02223       &     7.415x10$^{-4}$  \\            
10Ac  &         0.67670       &     2.257x10$^{-2}$  &          0.01699  &          5.667x10$^{-4}$  &   0.01658      &      5.531x10$^{-4}$  \\             
11Ac  &         0.61638       &     2.056x10$^{-2}$  &          0.01295  &          4.320x10$^{-4}$  &   0.01268      &      4.230x10$^{-4}$  \\             
12Ac  &         0.56597      &      1.888x10$^{-2}$  &          0.01009   &         3.366x10$^{-4}$  &   0.00992      &      3.309x10$^{-4}$  \\             
13Ac  &         0.52320      &      1.745x10$^{-2}$  &          0.00802   &         2.675x10$^{-4}$  &   0.00790      &      2.635x10$^{-4}$  \\             
14Ac  &         0.48644       &     1.623x10$^{-2}$  &          0.00648   &         2.161x10$^{-4}$  &   0.00639      &      2.131x10$^{-4}$  \\             
15Ac  &         0.45452       &     1.516x10$^{-2}$  &          0.00531  &          1.771x10$^{-4}$  &   0.00524      &      1.748x10$^{-4}$  \\             
16Ac  &         0.42653       &     1.423x10$^{-2}$  &          0.00440   &         1.468x10$^{-4}$  &   0.00436      &      1.454x10$^{-4}$  \\             
17Ac  &         0.40180       &     1.340x10$^{-2}$  &          0.00369   &         1.231x10$^{-4}$  &   0.00366      &     1.221x10$^{-4}$  \\             
18Ac  &         0.37978       &     1.267x10$^{-2}$  &          0.00313     &       1.044x10$^{-4}$  &   0.00310      &      1.034x10$^{-4}$  \\             
19Ac  &         0.36004       &     1.201x10$^{-2}$  &          0.00267    &        0.891x10$^{-4}$  &   0.00265      &      0.884x10$^{-4}$  \\             
20Ac  &         0.34225       &     1.142x10$^{-2}$  &          0.00230    &        0.767x10$^{-4}$  &   0.00228      &     0.761x10$^{-4}$  \\  
      \noalign{\smallskip}
            \noalign{\smallskip}
            \hline
           \end{tabular}
{\small \begin{list}{}{} 
\item[]   $^a$ Pentacene, experimental values from Heinecke et al. (1998).
\end{list}}
\end{table*}

In interstellar conditions, the rotational broadening of the expected profiles of absorption bands of polyacenes may be roughly estimated as follows. As quoted in Section 2.1, typical values for the rotation constants, for example\ for N\,=\,14,  are A\,$\sim$\,1.6x10$^{-2}$\,cm$^{-1}$ ($\sim$\,2.3x10$^{-2}$\,K), B\,$\sim$\,2.2x10$^{-4}$\,cm$^{-1}$ and C\,$\sim$\,2.1x10$^{-4}$\,cm$^{-1}$, that is,\  A/(B or C)\,$\sim$\,75. We may therefore neglect the terms depending on B or C in the rotational energy, and assume the molecule to be a symmetric top with rotational energy, 

\begin{equation}
{\rm E_{rot}   \approx A \times K^2 \approx  0.023 \times K^2 ~~Kelvin}
.\end{equation}

Therefore, at rotational temperature T$_{\rm rot}$, a typical value of K is

\begin{equation}
{\rm K_m = \sqrt{44xT_{rot}} = 67  \sqrt{T_{rot}/100}}
.\end{equation}

Rotational broadening occurs because of the variation of the energy difference between the ground and the excited levels $f$ and $e$ of the molecule. For a transition moment polarized along the long axis of the molecule, as expected for the main D$_2$ band of polyacenes (Fig.\ 2), K$_e$\,=\,K$_f$\,=\,K.  The shift of the K line is thus such that 

\begin{equation}
{\rm  \Delta \nu = (A}_e - {\rm A}_f) \times {\rm K^2} = \delta {\rm A \times K^2}
.\end{equation}

Writing A$_f$/$\delta$A = 100\,$p$ yields a typical shift for K\,=\,K$_{\rm m}$ and N\,=\,14,

\begin{equation}
{\rm  \Delta \nu_m} = 0.7 \times p \times ({\rm T_{rot}/100) ~cm^{-1}} 
.\end{equation}

As we may expect that the spread in $\Delta \nu$, that is,\ the width of the DIB, may reach up to 2$\times {\rm  \Delta \nu_m}$, 
values of $p$\,$\sim$\,1-2, or (A$_e$-A$_f$)/A$_f$\,$\sim$\,1-2\%, and ${\rm T_{rot}}$\,$\sim$\,100\,K (in agreement with values of Appendix B) might explain the observed widths of $\zeta$ DIBs $\sim$2.0-2.5\,cm$^{-1}$ in normal lines of sight (e.g., Hobbs et al.\ 2008, 2009). Such values of  (A$_e$-A$_f$)/A$_f$ do not appear unlikely, but they should be proved by detailed calculations for such polyacenes. 

We note that for the case of the B$_{2u}$ band of pentacene measured by Heinecke et al.\ (1998), A$_e$-A$_f$ has a reverse sign with (A$_e$-A$_f$)/A$_f$\,$\sim$\,-1\%. However, this is a band with a different symmetry with its transition moment perpendicular to  the long axis of the molecule. We also note that its large broadening at low temperature is due to the transitions with $\Delta$K\,=\,$\pm$1 for such a symmetry.

\section{Computational details}

The computations employed Becke’s (Becke 1993) three parameter hybrid functional in conjunction with the correlation functional of Lee, Yang, and Parr (Lee et al\, 1988) as implemented (Stephens et al.\ 1994) in the Gaussian 09 program (Frisch et al\, 2009). The spin-unrestricted ansatz was employed for the neutral acenes as well as for any acene radicals. The 6-31G* basis set was employed (Hehre et al.\ 1972). The geometries of all stationary points were fully optimized at the UB3LYP/6-31G* level of theory and harmonic vibrational frequencies were obtained analytically confirming that the geometries obtained correspond to minima on their respective potential energy surface.

\bigskip

{\bf References}

Ali-Ha{\"i}moud, Y., Hirata, C.~M., \& Dickinson, C.\ 2009, MNRAS, 395, 1055

Ali-Ha{\"i}moud, Y.\ 2013, Advances in Astronomy, 2013, 462697 

Allamandola, L.~J., Tielens, A.~G.~G.~M., \& Barker, J.~R.\ 1985, \apjl, 290, L25 

Allamandola, L.~J., Hudgins, D.~M., Bauschlicher, C.~W., Jr., \& Langhoff, S.~R.\ 1999, \aap, 352, 6

Andrews, H., Boersma, C., Werner, M.~W.\ et al.\ 2015, \apj, 807, 99

Andrews, H., Candian, A., \& Tielens, A.~G.~G.~M.\ 2016, \aap, 595, A23 

Anthony, J.~E.\ 2008, Angew.\ Chem.\ Int.\ Ed., 47, 452

Battaglia, S., Faginas-Lago, N., Andrae, D., Evangelisti, S.\ \& Leininger, T.\ 2017, 
J.\ Phys.\ Chem.\ A, 121, 3746

Becke, A.~D.\ 1993 J. Chem.\ Phys., 98, 5648

Bettinger, H.~F.\ \&  T\"{o}nshoff, C.\ 2015, Chem.\ Rec., 15, 364

Bettinger, H. F.,  T\"{o}nshoff, C., Doerr C.~M.\ \& Sanchez-Garcia E.\ 2016, J. Chem. Theory Comput. 12, 305

Cami, J., Salama, F., Jim{\'e}nez-Vicente, J., Galazutdinov, G.~A., \& Kre{\l}owski, J.\ 2004, \apjl, 611, L113 

Cami, J.\ 2014,  The Diffuse Interstellar Bands, IAU Symposium, 297, 370 

Cami, J., \& Cox, N.~L.~J.\ 2014,  The Diffuse Interstellar Bands, IAU Symposium, 297

Cami, J., Cox, N.~L., Farhang, A., Smoker, J., Elyajouri, M., Lallement, R.\ et al.\ 2018, The Messenger, 171, 31 

Campbell, E.~K., Holz, M., Gerlich, D., \& Maier, J.~P.\ 2015, \nat, 523, 322 

Campbell, E.~K., Holz, M., Maier, J.~P., et al.\ 2016a, \apj, 822, 17

Campbell, E.~K., Holz, M., \& Maier, J.~P.\ 2016b, \apjl, 826, L4 

Campbell, E.~K. \& Maier, J.~P., 2017a, J.\ Chem.\ Phys., 146, 160901

Campbell, E.~K., \& Maier, J.~P.\ 2017b, \apj, 850, 69 

Campbell, E.~K., \& Maier, J.~P.\ 2018, \apj, 858, 36 

Castellanos, P., Candian, A., Andrews, H., \& Tielens, A.~G.~G.~M.\ 2018a, \aap, 616, A166 

Castellanos, P., Candian, A., Zhen, J., Linnartz, H., \& Tielens, A.~G.~G.~M.\ 2018b, \aap, 616, A167  

 Cernicharo, J., Heras, A.~M., Tielens, A.~G.~G.~M., et al.\ 2001, \apjl, 546, L123

Chabot, M., B{\'e}roff, K., Dartois, E., Pino, T., \& Godard, M.\ 2017, arXiv:1709.07803 

Chai, J.~D.\ 2014, J.\ Chem.\ Phys., 140, 18A521

Chiar, J.~E., Tielens, A.~G.~G.~M., Adamson, A.~J., \& Ricca, A.\ 2013, \apj, 770, 78 

 Cocchi, C., Prezzi, D., Ruini, A., Caldas, M.~J., \& Molinari, E.\ 2014, Journal of Physical Chemistry A, 118, 6507 

Crawford, M.~K., Tielens, A.~G.~G.~M., \& Allamandola, L.~J.\ 1985, \apjl, 293, L45 

Dahlstrom, J., York, D.~G., Welty, D.~E., et al.\ 2013, \apj, 773, 41

Douglas, A.~E.\ 1977, Nature, 269, 130 

Duley, W.~W.\ 2000, \apj, 528, 841 

Elyajouri, M., Lallement, R., Cox, N.~L.~J., et al.\ 2018, A\&A, \aap, 616, A143 

Ensor, T., Cami, J., Bhatt, N.~H., \& Soddu, A.\ 2017, \apj, 836, 162 

Fan, H., Welty, D.~E., York, D.~G., et al.\ 2017, \apj, 850, 194 

Foing, B.~H., \& Ehrenfreund, P.\ 1994, \nat, 369, 296

 Frisch, M.~J., Trucks, G.~W., Schlegel, H.~B.\ et al.\ 2009,  Gaussian 09, Revision A.02, Wallingford CT, 2009 

Gredel, R., Carpentier, Y., Rouill{\'e}, G., et al.\ 2011, \aap, 530, A26 

Hachmann, J., Dorando, J.~J., Avil\'es, M., \& Garnet  Chan, G.~K.\ 2007, J.\ Chem.\ Phys., 127, 134309

Halasinski, T.~M., Hudgins, D.~M., Salama, F., \& Allamandola, L.~J.\ 2000, J.\ Phys.\ Chem.\ A, 104, 7484

Halasinski, T.~M.,  Weisman, J.~L.,  Ruiterkamp, R.\ et al.\ 2003, J.\ Phys.\ Chem.\ A, 107, 3660

Hahndorf, I., Lee, Y.~T.,  Kaiser, R.~I., Vereecken, L., Peeters, J., Bettinger, H.~F.\ et al.\ 2002, J.\ Chem.\ Phys., 116, 3248

Hardy, F.-X., Rice, C.~A., \& Maier, J.~P.\ 2017, \apj, 836, 37 

Hehre, W.~J., Ditchfield, R.\ \& Pople, J.~A.\ 1972, J.\ Chem.\ Phys., 56, 2257

Heinecke, E., Hartmann, D., M\"{u}ller, R.\ \& Hese, A.\ 1998, J.\ Chem.\ Phys., 109, 906

Herbig, G.~H.\ 1995, \araa, 33, 19 

Hoang, T., Draine, B.~T., \& Lazarian, A.\ 2010, \apj, 715, 1462 

Hoang, T., Lazarian, A., \& Draine, B.~T.\ 2011, \apj, 741, 87 

Hobbs, L.~M., York, D.~G., Snow, T.~P., et al.\ 2008, \apj, 680, 1256 

Hobbs, L.~M., York, D.~G., Thorburn, J.~A., et al.\ 2009, \apj, 705, 32 



Huang, J., \& Oka, T.\ 2015, Molecular Physics, 113, 2159 

Huang, R., Phan, H., Herng, T.~S.\ et al.\ 2016, J.\ Am.\ Chem.\ Soc., 138, 10323

Huisken, F., Rouill{\'e}, G., Steglich, M., et al.\ 2014,  The Diffuse Interstellar Bands,  IAU Symposium, 297, 265  

Ji, M., Bernard, J., Chen, L.\ et al.\ 2017, J.\ Chem.\ Phys., 146, 044281

Jones, A.~P., Ysard, N., K{\"o}hler, M., et al.\ 2014, Faraday Discussions, 168, 313 

Jones, A.~P.\ 2016, Royal Society Open Science, 3, 160223 

Kaiser, R.~I.,  Hahndorf, I., Lee, Y.~T., Huang, H.~C.~L., Bettinger, H.~F.\ et al.\ 1999, J.\ Chem.\ Phys., 110, 6091

Kaiser, R.~I., Vereecken, L., Peeters, J., Bettinger, H.~F.\ et al.\ 2003, \aap, 406, 385

Ka{\'z}mierczak, M., Schmidt, M.~R., Bondar, A., \& Kre{\l}owski, J.\ 2010, \mnras, 402, 2548 

Kofman, V., Sarre, P.~J., Hibbins, R.~E., ten Kate, I.~L., \& Linnartz, H.\ 2017, Molecular Astrophysics, 7, 19

Krasnokutski, S.~A., Huisken, F., J{\"a}ger, C., \& Henning, T.\ 2017, \apj, 836, 32 

Kroto, H.~W., \& Jura, M.\ 1992, \aap, 263, 275

Kr{\"u}ger, J., Giarca, F., Eisenhut, F.\ et al.\ 2017, Angew.\ Chem.\ Int.\ Ed., 56, 11945

 Lallement, R., Cox, N.~L.~J., Cami, J., et al.\ 2018, \aap, 614, A28 

Lan, T.-W., Ménard, B., \& Zhu, G.\ 2015, MNRAS, 452, 3629 

Leach, S.\ 1995, \planss, 43, 1153 

Leach, S.\ 1996, Z. Physikalische Chemie, 195, 15

 Lee, C.\, Yang,W.\ \& Parr, R.~G.\ 1988, Phys.\ Rev.\ B, 37, 785

Leger, A., \& Puget, J.~L.\ 1984, \aap, 137, L5 

L\'eger, A., \& D'Hendecourt, L.\ 1985, \aap, 146, 81 

Leger, A., D'Hendecourt, L., \& Boissel, P.\ 1988, Physical Review Letters, 60, 921

Le Page, V., Snow, T.~P., \& Bierbaum, V.~M.\ 2001, \apjs, 132, 233 

Le Page, V., Snow, T.~P., \& Bierbaum, V.~M.\ 2003, \apj, 584, 316 

Li, J.\ \& Zhang, Q.\ 2015, Appl.\ Mater.\ Interfaces, 7, 28049

Li, J., Chen, S., Wang, Z.\ \& Zhang, Q.\ 2016, Chem. Rec., 16, 1518

Luzanov,A.~V.,  Plasser, F., Das, A.\ \& Lischka, H.\ 2017, J.\ Chem.\  Phys., 146, 064106

Mackie, C.~J., Peeters, E., Bauschlicher, C.~W., Jr., \& Cami, J.\ 2015, \apj, 799, 131 

Maier, J.~P., Walker, G.~A.~H., \& Bohlender, D.~A.\ 2004, \apj, 602, 286

Malloci, G., Mulas, G., Cappellini, G., \& Joblin, C.\ 2007, Chemical Physics, 340, 43 

Malloci, G., Cappellini, G., Mulas, G.\ \& Mattoni, A.\ 2011, Chemical Physics, 384, 19

Marshall, C.~C.~M., Kre{\l}owski, J., \& Sarre, P.~J.\ 2015, \mnras, 453, 3912 

McGuire, B.~A., Burkhardt, A.~M., Kalenskii, S., et al.\ 2018, Science, 359, 202 

Micelotta, E.~R., Jones, A.~P., \& Tielens, A.~G.~G.~M.\ 2010, \aap, 510, A37
 
Mondal, R., T\"{o}nshoff, C.,  Khon, D.\ et al.\ 2009, J.\ Am.\ Chem.\ Soc.,  131, 14281

Montillaud, J., Joblin, C., \& Toublanc, D.\ 2013, \aap, 552, A15 

Nijegorodov, N., Ramachandran,  V.\ \& Winkoun, D.~P.\ 1997, Spectrochimica Acta  A, 53, 1813

Oka, T., Welty, D.~E., Johnson, S., et al.\ 2013, \apj, 773, 42; 2014, \apj, 793, 68

Omont, A.\ 2016, \aap, 590, A52 


Parneix, P., Gamboa, A., Falvo, C., et al.\ 2017, Molecular Astrophysics, 7, 9

Pendleton, Y.~J., \& Allamandola, L.~J.\ 2002, \apjs, 138, 75  

Planck Collaboration, Ade, P.~A.~R., Aghanim, N., et al.\ 2014, \aap, 565, A103 

Plasser, F., Pas\v{a}li\'c, H., Gerzabek, M.~H., et al.\ 2013, Angew.\ Chem.\ Int.\ Ed., 52, 2581

Povie, G., Segawa, Y., Nishihara, T., Miyauchi, Y.\ \&  Itami, K.\ 2017, Science, 356, 172

Povie, G., Segawa, Y., Nishihara, T., Miyauchi, Y.\ \&  Itami, K.\ 2018, J.\ Am.\ Chem.\ Soc.,  140, 10054

Puget, J.~L., \& Leger, A.\ 1989, \araa, 27, 161 

Rai, D.~K.\ \& Shukla, K.\ 2018, arXiv:1810.03482v2

Rouan, D., Leger, A., Omont, A., \& Giard, M.\ 1992, \aap, 253, 498 

Rouan, D., Leger, A., \& Le Coupanec, P.\ 1997, \aap, 324, 661 

Ruiterkamp, R., Halasinski, T., Salama, F., et al.\ 2002, \aap, 390, 1153 

 Ruiterkamp, R., Cox, N.~L.~J., Spaans, M., et al.\ 2005, \aap, 432, 515 

Salama, F., Bakes, E.~L.~O., Allamandola, L.~J., \& Tielens, A.~G.~G.~M.\ 1996, \apj, 458, 621 

Salama, F., Galazutdinov, G.~A., Kre{\l}owski, J., et al.\ 2011, \apj, 728, 154 

Salama, F., \& Ehrenfreund, P.\ 2014,  The Diffuse Interstellar Bands, IAU Symposium, 297, 364 

Sarre, P.~J.\ 2006, Journal of Molecular Spectroscopy, 238, 1 

Sarre, P.~J.\ 2014, The Diffuse Interstellar Bands, IAU Symposium, 297, 34 



Scott, A., Duley, W.~W., \& Pinho, G.~P.\ 1997, \apjl, 489, L193 


Shen, B., Tatchen, J., Sanchez-Garcia, E.\ \& Bettinger, H.~F.\  2018, Angew. Chem. Int. Ed. 2018, 

 Silsbee, K., Ali-Ha{\"i}moud, Y., \& Hirata, C.~M.\ 2011, \mnras, 411, 2750

Simon, A., Champeaux, J.~P., Rapacioli, M.\ et al.\ 2018, Theoretical Chemistry accounts, 137:106

Snow, T.~P.\ 1995, in Tielens, A.~G.~G.~M., \& Snow, T.~P., The Diffuse Interstellar Bands, {\it Astrophysics and Space Science Library}, 202, Kluwer

Snow, T.~P.\ 2002, \apj, 567, 407 

Snow, T.~P.\ 2014,  The Diffuse Interstellar Bands, IAU Symposium, 297, 3 



Spieler, S., Kuhn, M., Postler, J., et al.\ 2017, \apj, 846, 168 

Steglich, M., Bouwman, J., Huisken, F., \& Henning, T.\ 2011, \apj, 742, 2  

StephensT, P. J., Devlin, F.~J., Chabalowski, C.~F.\ \& Frisch, M.~J.\ 1994, J.\ Phys. Chem., 98, 11623

Thaddeus, P.\ 1995, in Tielens, A.~G.~G.~M., \& Snow, T.~P.\ 1995, The Diffuse Interstellar Bands, {\it Astrophysics and Space Science Library}, 202, Kluwer

Thorburn, J.~A., Hobbs, L.~M., McCall, B.~J., et al.\ 2003, \apj, 584, 339

Tian, M., Liu, B.~S., Hammonds, M., Wang, N., Sarre, P.~J.\ \& Cheung, A.~S.-C.\ 2012, Phys.\ Chem.\ Chem.\ Phys., 14, 6603

Tielens, A.~G.~G.~M., \& Snow, T.~P., 1995. The Diffuse Interstellar Bands, {\it Astrophysics and Space Science Library}, 202, Kluwer

Tielens, A.~G.~G.~M.\ 2008, \araa, 46, 289 

Tielens, A.~G.~G.~M.\ 2013, Reviews of Modern Physics, 85, 1021

T{\"o}nshoff \& Bettinger 2010, Angew.\ Chem., Int.\ Ed., 49, 4125

van der Zwet, G.~P., \& Allamandola, L.~J.\ 1985, \aap, 146, 76 

Vos, D.~A.~I., Cox, N.~L.~J., Kaper, L., Spaans, M., \& Ehrenfreund, P.\ 2011, \aap, 533, A129 

Walker, G., Bohlender, D., Maier, J., \& Campbell, E.\ 2015, \apjl, 812, L8 

 Walker, G.~A.~H., Campbell, E.~K., Maier, J.~P., Bohlender, D., \& Malo, L.\ 2016, \apj, 831, 130 

Walker, G.~A.~H., Campbell, E.~K., Maier, J.~P., \& Bohlender, D.\ 2017, \apj, 843, 56 

West, B., Useli-Bacchitta, F., Sabbah, H., et al.\ 2014, J.\ Phys.\ Chem.\ A, 118, 7824

Wu, C., Lee, P.\ \& Chai, J.\ 2016, Scientific Reports, 6, 37249

Yang, Y., Davidson, E.~R.\ \& Yang, W.\ 2016,  Proc.\ Natl.\ Acad.\ Sci.\ USA, 113, E5098-E510721113

Ye, Q.\ \& Chi, C.\ 2014, Chem.\ Mater., 26, 4046

Ysard, N., \& Verstraete, L.\ 2010, \aap, 509, A12 

Zack, L.~N., \& Maier, J.~P.\ 2014,  The Diffuse Interstellar Bands,  IAU Symposium, 297, 237

Zade, S.~S.\  \& Bendikov, M.\ 2012, J.\ Phys.\ Org.\ Chem., 25, 452

Zhao, T.~Q., Li, Q., Liu, B.~S., Gover, R.~K.~E., Sarre, P.~J.\ \& Cheung, A.~S.-C.\ 2016, Phys.\ Chem.\ Chem.\ Phys., 18, 3489

Zhen, J., Castellanos, P., Paardekooper, D.~M., Linnartz, H., \& Tielens, A.~G.~G.~M.\ 2014, \apjl, 797, L30 

Zhen, J., Castellanos, P., Paardekooper, D.~M., et al.\ 2015, \apjl, 804, L7

Zhen, J., Rodriguez Castillo, S., Joblin, C., et al.\ 2016, \apj, 822, 113 

Zhen, J., Candian, A., Castellanos, P., et al.\ 2018, \apj, 854, 27  

Zuzak,R., Dorel, R., Krawiec, M.\ et al.\ 2017, ACS Nano, 11, 9321

Zuzak, R., Dorel, R.,  Kolmer, M., Szymonski, M., Godlewski, S.\ \& Echavarren, A. M.\ 2018, Angew. Chem. Int. Ed. 2018, 57, 10500

\end{document}